\documentclass[twocolumn,floatfix,superscriptaddress,longbibliography,nofootinbib]{revtex4-1}
\RequirePackage[sort&compress]{natbib}

\usepackage{color}

\usepackage{amssymb,amsfonts,amsmath}
\usepackage{bm}
\usepackage[pdftex]{graphicx}
\usepackage{xcolor}
\usepackage{subfigure}
\usepackage{comment}

\usepackage{hyperref}
\hypersetup{
    colorlinks=true,
    linkcolor=blue,
    filecolor=magenta,      
    urlcolor=blue,
}
 
\definecolor{darkGray}{RGB}{153,153,153}
\definecolor{darkBlue}{RGB}{37,113,161}
\definecolor{darkGreen}{RGB}{113,161,37}
\definecolor{darkRed}{RGB}{186,21,24}

\newcommand{\rev}[1]{\textcolor{black}{#1}}

\begin{document}

\title{Modeling Structure and Resilience of the Dark Network}

\author{Manlio De Domenico}\thanks{To whom correspondence should be addressed; E-mail:  manlio.dedomenico@urv.cat.}
\affiliation{Departament d'Enginyeria Inform\`{a}tica i Matem\`{a}tiques, Universitat Rovira i Virgili, 43007 Tarragona, Spain}

\author{Alex Arenas}
\affiliation{Departament d'Enginyeria Inform\`{a}tica i Matem\`{a}tiques, Universitat Rovira i Virgili, 43007 Tarragona, Spain}

\begin{abstract}
While the statistical and resilience properties of the Internet are no more changing significantly across time, the Darknet, a network devoted to keep anonymous its traffic, still experiences rapid changes to improve the security of its users. Here, we study the structure of the Darknet and we find that its topology is rather peculiar, being characterized by non-homogenous distribution of connections -- typical of scale-free networks --, very short path lengths and high clustering -- typical of small-world networks -- and lack of a core of highly connected nodes. 

We propose a model to reproduce such features, demonstrating that the mechanisms used to improve cyber-security are responsible for the observed topology. Unexpectedly, we reveal that its peculiar structure makes the Darknet much more resilient than the Internet -- used as a benchmark for comparison at a descriptive level -- to random failures, targeted attacks and cascade failures, as a result of adaptive changes in response to the attempts of dismantling the network across time. 
\end{abstract}

\maketitle

\section{Introduction}

Since when Internet became a publicly accessible infrastructure and communication network, its resilience to random failures -- caused, for instance, by unexpected crashes due to nodes' malfunction or protocol's errors -- or attack -- actions devoted to isolate nodes that play a vital role in the network -- has been widely investigated\cite{albert2000error,cohen2000resilience,callaway2000network,pastor2001dynamical,pastor2007evolution}. In fact, the Internet exhibits highly nontrivial structural and dynamical properties, from heavy-tail distribution of connections -- known as scale-free property\cite{barabasi1999emergence} -- to a moderate amount of clustering -- proportional to the fraction of nodes that form closed triangles --, whose modeling has been the subject of intense research activity\cite{vazquez2003topology,serrano2005competition,serrano2006modeling,boguna2010sustaining,papadopoulos2012popularity}. In fact, several years after the Internet first proper crash, in 1980, the focus of many studies has been, and still is, to improve its resilience\cite{savage1999detour,fortz2004increasing,doyle2005robust,boguna2009navigability,boguna2009navigating,boguna2010sustaining,briscoe2014byte}. 
In late 90s, about 30 years after the first Internet prototype, the US Defense Advanced Research Projects Agency (DARPA) and the Office of Naval Research started to develop a communication network, at the application layer, based on anonymous connections and, in principle, resistant to both eavesdropping and traffic analysis\cite{syverson1997private}. This network was based on \emph{onion routing}, a special infrastructure for private communications over a public network that is able to hide the content of a message and the identity of peers who are exchanging it\cite{goldschlag1999onion}. Nowadays, this infrastructure is better known as Tor network and represents the backbone of the Darknet, an Web of hidden services that are not reachable from within the Internet. The Darknet turned out to be the most suitable communication network to exchange sensitive information, both licit and illicit, becoming soon the target of governments trying to identify dissidents or of intelligence agencies, such as CIA and GCHQ\cite{guardian2013}, to contain unauthorized news leaks, distribution of illegal contents or trade of illegal substances.

Here, we characterize the structural properties of the Darknet across time, from 2013 to 2015, and we compare them against the Internet topology. It is worth remarking that throughout the manuscript we model and characterize the Darknet from a complex system perspective, meaning that we attach to the representation of both the Internet and the Darknet from available data, while focusing the study on the particular networks structures they provide. Note that this comparison is performed at a descriptive level, using the structure of the Internet as a benchmark to highlight the salient features of the Darknet. The Autonomous Systems data capture connections at the Internet layer (IP packages), while Tor works on the application layer, which means that is built on top of the Internet and transport layers.
We propose a model, based on how Tor functions, to reproduce with high accuracy the most salient characteristics of the Darknet.
Finally, we perform a thorough analysis, based on simulations, of the resilience of both networks to three different types of failures, static -- due to random disruptions or targeted attacks\cite{albert2000error} -- and dynamical -- due to the cascade failures induces by attacking a single specific node of the network\cite{motter2002cascade,motter2004cascade}, and show that the Darknet is much more robust than the Internet under any perspective.

%\section{Results}

\section{Overview of the data sets} For our analysis, we use publicly available data sets for both the Internet and the Darknet. The Internet topology, at the level of autonomous systems (AS), is sampled from historical AS-level topology data derived from Border Gateway Protocol (BGP) monthly snapshots, consisting of IPv4 and IPv6 links appearing between different endpoints during that month. The data are hosted by the UCLA Computer Science Department's Internet Research Lab\cite{internet-data}.

The Darknet topology is sampled from the data obtained by probing the Tor network to improve its performance\cite{annessinavigator}. The links between endpoints are extracted from the chain of circuits built by Tor clients to probe the network. The network is directed, but we will treat it as undirected, in the following. The full raw data are available upon request and a partial release can be downloaded from a public repository\cite{darknet-data}. \rev{Although such data were obtained to study the performance of the Tor network and not its topology, they provide the best approximation to the underlying topology of the Darknet to date.}

\begin{figure*}
\centering
\includegraphics[width=18cm]{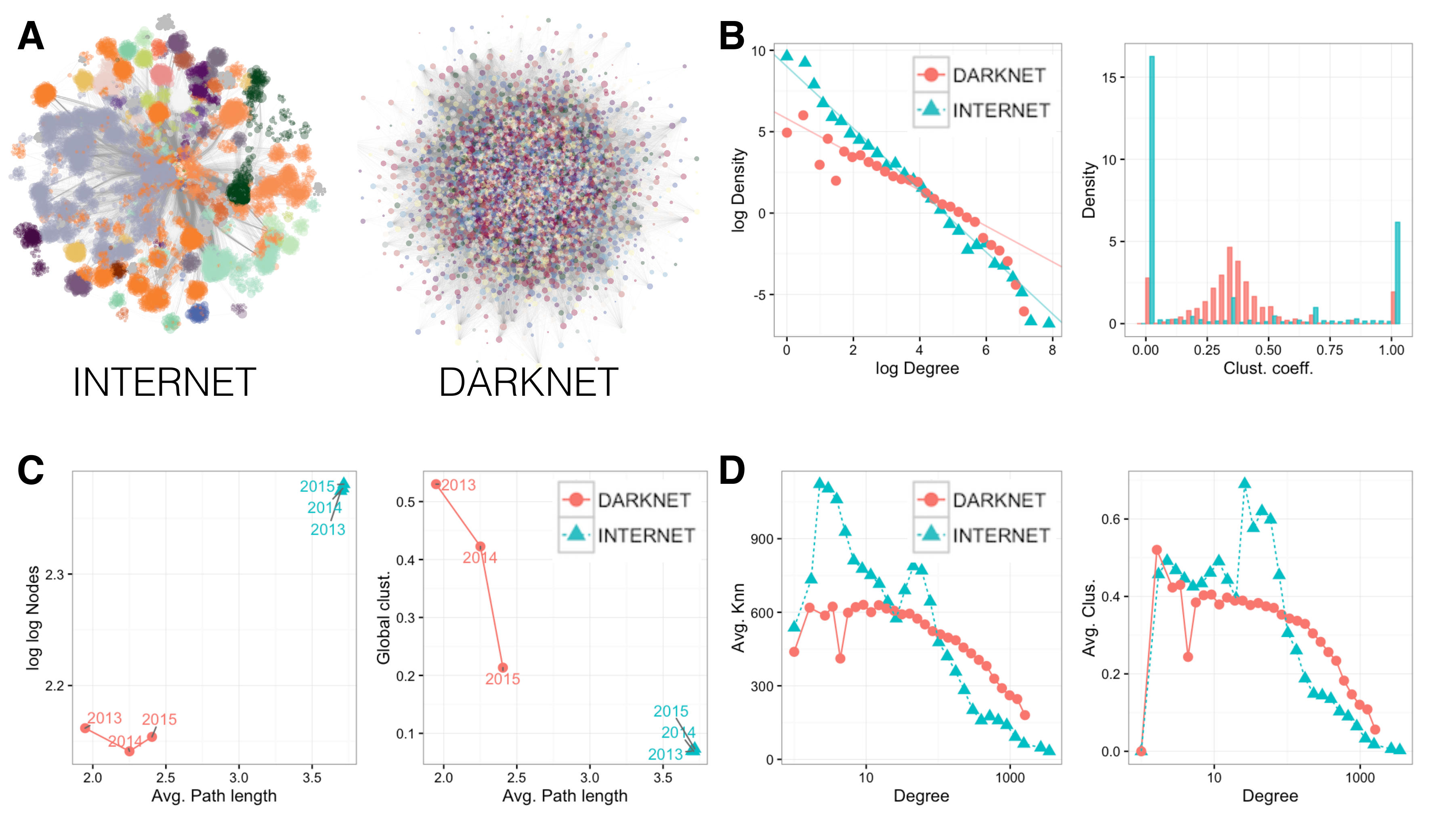}
\caption{{\bf Structural analysis of Internet and Darknet topologies.} Force-directed visualization of the ({\bf A}) Internet and the Darknet in 2015, with nodes colored to put in evidence the underlying mesoscale structure. ({\bf B}) Density of the degree (solid lines are for guidance only) and local cluster coefficient for the two networks in 2015. ({\bf C}) Scatter plot of network sizes, average path length and global clustering coefficient for the three temporal snapshots considered in this study. ({\bf D}) Average nearest-neighbors' degree (left) and average local clustering coefficient (right) against degree, to characterize higher-order correlations (see Supplementary Figure~1 for other structural descriptors and their evolution between 2013 and 2015).}
\label{fig:structural}
\end{figure*}

In both cases, we have been able to build three temporal snapshots, corresponding to the networks in December 2013, May 2014 and January 2015. The Internet network snapshots have 46,462, 47,626 and 49,635 nodes with 195,446, 204,254 and 221,470 connections, in the three periods, respectively. The Darknet network snapshots have 5,921, 4,953 and 5,535 nodes with 2,017,542, 536,287 and 274,831 connections, in the three periods, respectively. The structure of both networks for the 2015 period is shown in Fig.\ref{fig:structural}A.

\section{The structure of the Darknet}
\subsection{Characterizing the Darknet topology} Let us indicate with $A_{ij}^{[\xi]}(t)$ the entries of the adjacency matrix of each network ($\xi=$~Internet and Darknet) at time $t$ ($t=2013, 2014$ and 2015), with value equal to one if $i$ and $j$ are connected, and zero otherwise (here $i,j=1,2,...,N^{[\xi]}(t)$, where $N$ indicates the number of nodes in the network and $E$ the number of edges).

For any node $i$ in each network, we calculate the degree $k_{i}^{[\xi]}(t)=\sum\limits_{j}A^{[\xi]}_{ij}(t)$ -- characterizing the number of connections of each node -- and the local clustering coefficient $c_{i}^{[\xi]}(t)$ -- characterizing the tendency of nodes to form triangles --, defined by the ratio between the number of closed triangles involving node $i$ and the maximum number of triangles $\frac{1}{2}k_{i}^{[\xi]}(t)[k_{i}^{[\xi]}(t)-1]$ node $i$ might be part of. The mean degree is defined by $\bar{k}^{[\xi]}(t)=\langle k^{[\xi]}_{i}(t) \rangle$, whereas the average local clustering coefficient if given by $\bar{c}^{[\xi]}(t)=\langle c^{[\xi]}_{i}(t) \rangle$. Another macroscopic structural descriptor of interest is the global clustering coefficient $C^{[\xi]}(t)$, defined by the ratio between the total number of closed triplets and the total number of connected triplets of nodes in the network. In general, the values of $\bar{c}^{[\xi]}(t)$ and $C^{[\xi]}(t)$ are different for networks with a non-homogeneous connectivity. Throughout the analysis no power-law fitting is performed, the power laws indicated in the plots are a simple eye-guide.

The degree distribution is shown in Fig.~\ref{fig:structural} and exhibits highly inhomogeneity, with evident heavy tails that resemble two different truncated power laws with cut-off. The distribution of the local clustering coefficient is also different in the two cases, with much more unclustered nodes in the Internet than the Darknet. In the Internet, a significant fraction of nodes has local clustering equal to 1 and few nodes have intermediate values, at variance with the Darknet where the local clustering is more uniformly distributed and peaked around 0.3. On average, Darknet nodes have mean degree close to 100 and are more clustered than Internet nodes, which have mean degree close to 10. 

To quantify how easy is to transmit information between any two nodes of each network, we calculate the average path length $\ell^{[\xi]}(t)$, obtained by averaging the length of all shortest path connecting any pairs of nodes, and the diameter $D^{[\xi]}(t)$, defined as the length of the longest shortest path. The Internet has average path length close to 3.5, with diameter between 10 and 12, whereas the average path length in the Darknet ranges between 2 and 2.5, with diameter between 4 and 5 (see Supplementary Figure~1). Therefore, assuming no time constraints in the propagation of the information for both networks, communication in the Darknet is, in principle, much faster than the Internet, with the two most extremal nodes separated by no more than 5 hops, less than half of the Internet. Given that the traffic in the Darknet is encrypted and the routing is decentralized, at variance with the Internet, the shortest paths for communicating compensate the higher latency and throughput of the channels.

In Fig.~\ref{fig:structural}C is shown the relationship between the size of the networks, their average path length and their global clustering for the three snapshots under consideration. The presence of high clustering and short average path lengths are strong indicators that the two networks have nontrivial topology and formation mechanisms. In fact, both networks exhibit the property known as small-worldness\cite{watts1998collective}. Small-world networks are characterized by high clustering, with respect to random expectation, and characteristic length scaling as $\ell\sim \log N$. The average local clustering of the Internet is more than 1000 times higher than its uniformly random expectation, whereas the clustering of the Darknet is between 7 and 32 times larger than its uniformly random expectation, suggesting nontrivial triadic closure mechanisms underlying both networks. The characteristic length of the Darknet is larger than its uniformly random expectations, whereas this is not the case for the Internet. Although small-worldness is better understood as a tendency, rather than being quantified by a single number, it is worth remarking that $\ell(t)\approx\log N(t)$ -- as in small worlds -- for the Internet snapshots and $\ell(t)\approx\log \log N(t)$ -- as in ultra-small worlds\cite{cohen2003scale} -- for the Darknet ones.

The Internet and the Darknet also exhibit different types of higher-order correlations, as shown in Fig.~\ref{fig:structural}D, where the average nearest-neighbors' degree and the average local clustering coefficient are scattered against the degree. This kind of analysis is generally used to shade light on the degree-degree correlations of the network: if the degree or the clustering of each node are independent on the nodes in the neighborhood, no trends are expected. Instead, the two systems present highly anti-correlations, with hubs (i.e. nodes with largest degree) tending to be connected, on average, to nodes with much smaller degree and local clustering coefficient. This tendency is confirmed by the negative assortative mixing measured in both networks (see Supplementary Figure~1), defined by the Pearson's correlation coefficient of the degree of linked pairs of nodes\cite{newman2002assortative,newman2003mixing}.

From this structural analysis we find that while structural descriptors of the Internet do not change over time, this is not the case for the Darknet, that is still evolving (see Supplementary Figure~1 for other structural descriptors and their evolution between 2013 and 2015). Nevertheless, while the Internet has been widely investigated and several models have been proposed to explain its structure, the peculiar properties of the Darknet and the previous unavailability of data about its structure call for a model that is able to reproduce its most salient characteristics.

\subsection{Modeling the Darknet structure} To model the Darknet, it is crucial to understand how the Tor network functions. Any Tor client initially queries the Directory Authorities to get the consensus, a table providing information about all active nodes in the network and their metadata. The metadata is then evaluated to build a circuit, a chain of three nodes used to connect the client to the server where the (possibly hidden) service is located. The choice of the nodes of the circuit is subjected to severe constraints, by default. For instance, the same node can not be chosen twice and nodes run by the same operator are usually avoided. The choice of the first node is not performed uniformly random: instead, nodes with largest bandwidth are favored, with priority to long-lived nodes called ``guard''.

We model the above procedure with the following simple growing model. At time $\tau=0$ a small random network with $n_{0}\ll N$ nodes and $e_{0}$ edges is created first. We assign a timestamp $T(n)$ to such nodes ($n=1,2,...,n_{0}$), that will allow to calculate their age $\mathcal{A}(n, \tau)$ later at any time $\tau$, and a property $\mathcal{B}(n)$, not varying on time, whose value is sampled from a heavy-tailed distribution, to mimic the empirical bandwidth distribution. In the following, we will use a log-normal distribution.

\begin{figure}[!ht]
\centering
\includegraphics[width=6cm]{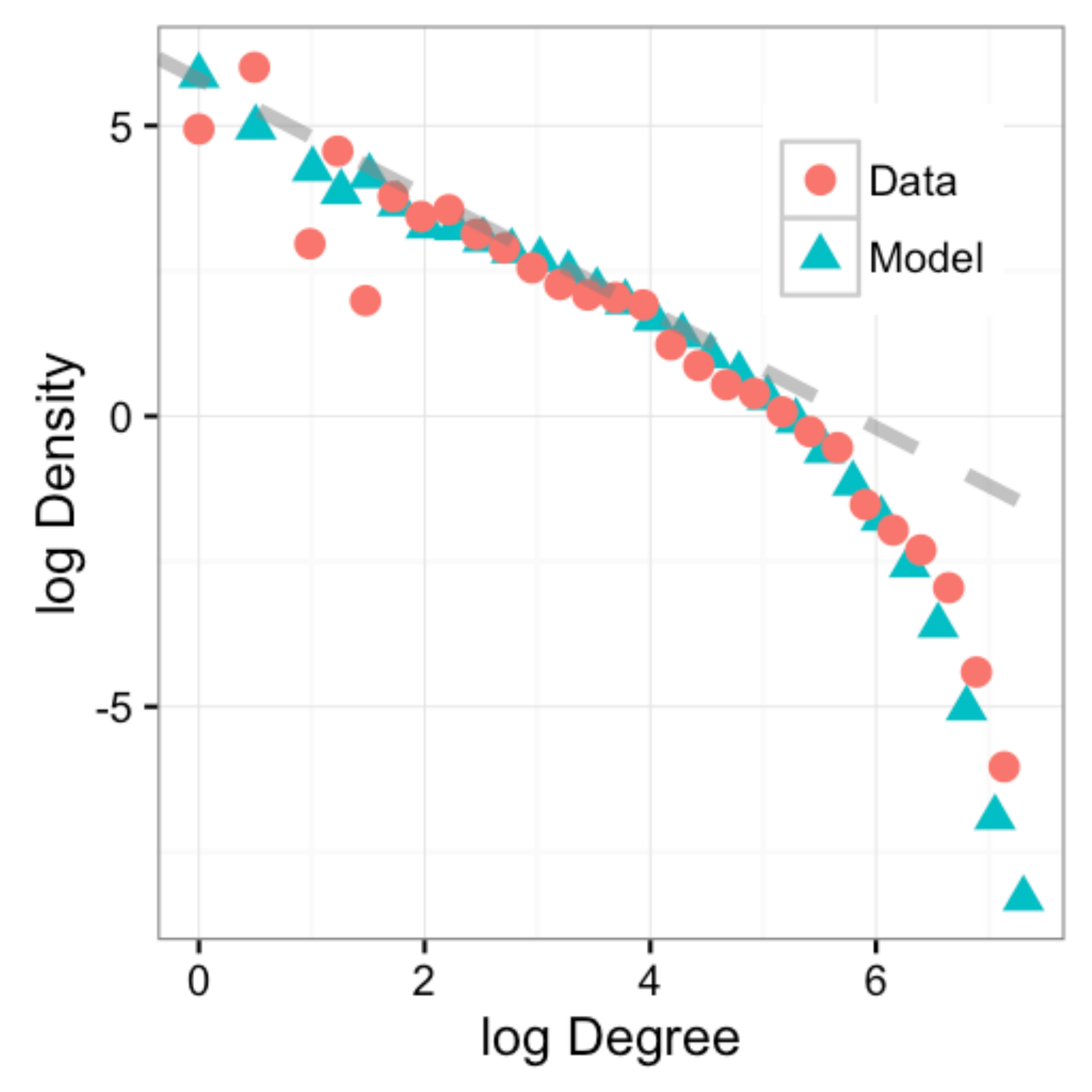}
\caption{{\bf Modeling the Darknet structure in 2015.} Degree distribution obtained from an ensemble of 50 random realizations of our model compared against the empirical distribution (the dashed line is for guidance only, to show the deviation of the degree distribution from a pure power-law with scaling exponent equal to -1.}
\label{fig:model-deg}
\end{figure}

At each time step $\tau=1,2,...,N-n_{0}$, a new node $n$ and $M$ links enter the network. The timestamp $T(n)=\tau$ and a bandwidth are assigned to $n$, as previously described. It is crucial to remark that the $M$ links do not involve node $n$ necessarily, at variance with processes based on the traditional preferential attachment. In fact, in a trust network like the Darknet, new nodes have to increase their reputation before being trusted and this is more likely to happen with aging. Nevertheless, links have to be created between trusted nodes at time $\tau$, therefore $2M$ nodes are randomly chosen with probability
\begin{eqnarray} 
p_{n'}(\tau) = \frac{\mathcal{A}^{\beta}(n',\tau)\mathcal{B}^{\gamma}(n')}{\sum\limits_{i=1}^{n}\mathcal{A}^{\beta}(i,\tau)\mathcal{B}^{\gamma}(i)},
\end{eqnarray}
where $\mathcal{A}(n',\tau)=\tau-T(n')$ is the age of node $n'$ at time $\tau$. The nodes are then randomly linked into $M$ pairs. Crucially, the degree of each node at each time step does not play any role in the growing process, that is completely driven by exogenous node's properties such aging and bandwidth. The final step is to rewire nodes randomly, while preserving the degree distribution: this last stage destroys possible structural correlations due to the previous stages of the model and it is crucial to introduce a higher level of randomness in connectivity. 

\begin{figure*}[!t]
\centering
\includegraphics[width=16cm]{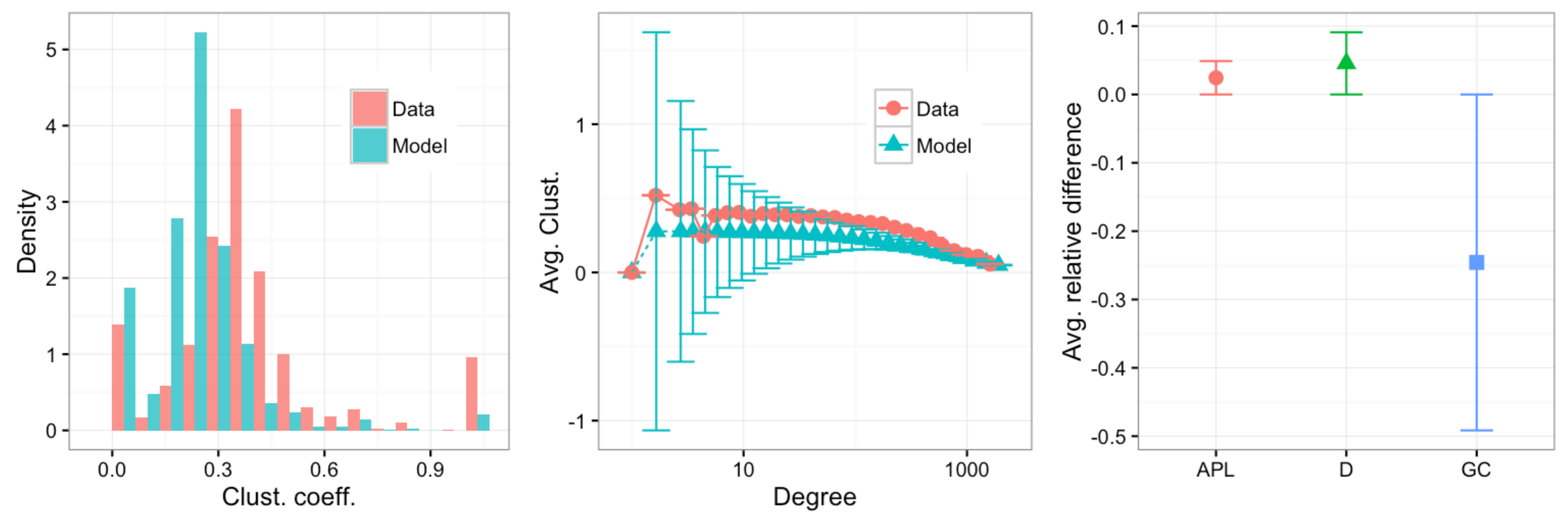}
\caption{{\bf Reproducing the Darknet (2015) structural descriptors.} Different structural descriptors obtained from an ensemble of 400 random realizations of our model compared against the observed values. In the last panel, we report the mean relative difference between the value calculated from the data and the values calculated from the model (``D'' is for diameter, ``APL'' is for average path length and ``GC'' is for global clustering).}
\label{fig:model-descr}
\end{figure*}

This Darknet stochastic model is very general and, varying the exponents $\beta$ and $\gamma$, it is possible to explore different scenarios, e.g. where probability is inversely proportional to age ($\beta<0$) and proportional to bandwidth ($\gamma>0$). In practice, the value of $M$ is fixed by the data as $M=(E - e_{0})/(N - n_{0})$ and therefore, the only free parameters of the model are, in general, $\beta$ and $\gamma$. In the following, we do not fit the parameters and we just consider the simplest scenario with linear proportionality ($\beta=\gamma=1$).

Our simulations revealed that this model generates networks that are remarkably close to the observed one from different perspectives. The high clustering and small characteristic length are satisfactorily reproduced, although the main finding here is that the degree distribution of simulated networks reproduce with excellent accuracy the empirical one, as shown in Fig.\,\ref{fig:model-deg}.

This result is of particular interest because, in general, networks with degree distribution scaling as power laws with exponent smaller than -2 (and especially close to -1) and cutoff are very difficult to model. Mechanisms involving degree-based preferential attachment and the influence of some exogenous properties, like aging and cost, have been proposed\cite{amaral2000classes} although they are able to reproduce power-law scaling with exponent equal or larger than -2 with cutoff.

The ensemble of random realizations of our model is sufficient to reproduce the main topological properties of the Darknet, including its structural resilience as we will see later. Figure\,\ref{fig:model-descr} shows the comparison between observed values and their expectation according to our model.

\subsection{Lack of ``rich club'' effect in the Darknet} 
\begin{figure}[!hb]
\centering
\includegraphics[width=6cm]{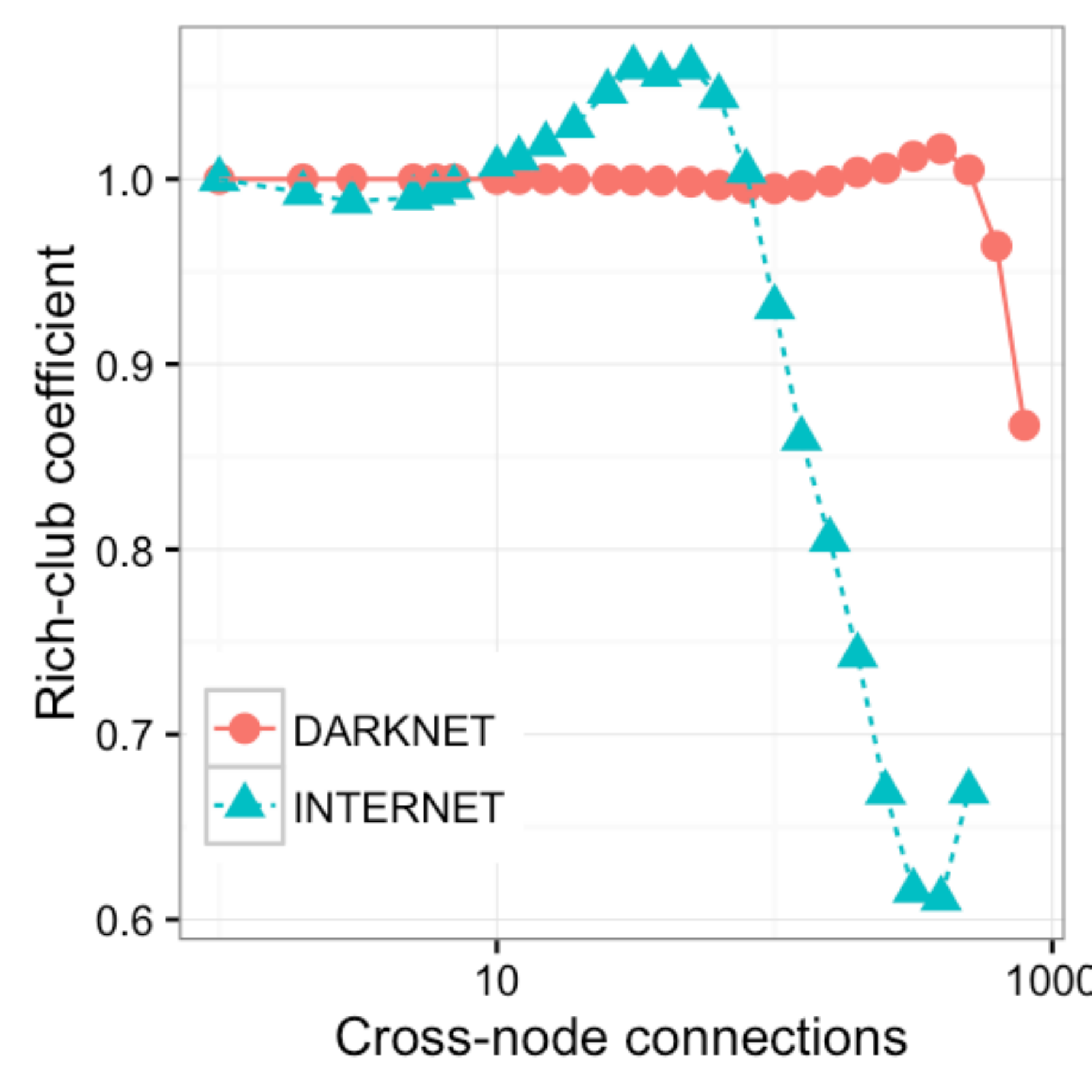}
\caption{{\bf ``Rich club'' structure of Internet and Darknet in 2015.} Ratio between the rich-club coefficient calculated for the empirical networks and its random expectation (see the text for further details) as a function of the degree threshold. Coefficient not shown for degree values for which the number of nodes is smaller than 100 (about 2\% of Darknet and 0.2\% of Internet).}
\label{fig:rich-club}
\end{figure}

It has been shown that in many complex networks, especially the Internet, nodes that are very central tend to interconnect more each other. This effect, called rich club, produces a core of nodes that is really important for the stability and the robustness of the network and can be quantified\cite{colizza2006detecting,opsahl2008prominence}. Let us denote by $E^{[\xi]}_{>k}(t)$ the number of connections among the $N^{[\xi]}_{>k}(t)$ nodes with degree larger than the threshold $k$. The rich-club coefficient is defined by
\begin{eqnarray}
\phi^{[\xi]}_{k}(t)=\frac{2E^{[\xi]}_{>k}(t)}{N^{[\xi]}_{>k}(t)[N^{[\xi]}_{>k}(t)-1]}.
\end{eqnarray}
However, to understand to which extent the observed rich-club effect is not due to chance, we generate an ensemble of $1000$ random networks preserving the empirical degree distribution and calculate the expected coefficient $\tilde{\phi}^{[\xi]}_{k}(t)$. Therefore we study how the ratio $\phi^{[\xi]}_{k}(t)/\tilde{\phi}^{[\xi]}_{k}(t)$ changes as a function of $k$: when this ratio is close to 1 the observed rich-club is compatible with random fluctuations, whereas when it is larger (smaller) than 1 it indicates the existence (absence) of a rich core of nodes. We show in Fig.~\ref{fig:rich-club} the calculated value of the ratio for the Internet and the Darknet in 2015. The Internet exhibits a clear rich-club effect for nodes with intermediate degree, around $k=50$, with largest hubs tending to be not interconnected each other. Conversely, the Darknet does not exhibit a rich core, with slight tendency of the largest hubs to be not interconnected each other, effect that in is magnitude significantly smaller than the case of the Internet.

Summing up, the Internet consists of a backbone of high-centrality nodes, whereas the Darknet does not. This result is compatible with the fact that nowadays Internet is a very centralized network providing an easier way to manage and search for online services, whereas the Darknet is very decentralized (as the Usenet, the ancestor of the Internet) but it is more difficult to manage and search for hidden online services.

\section{Resilience of the Darknet}

\subsection{Resilience to static failures} 

Here we investigate how the structural properties of the Darknet and the Internet are reflected in their resilience to perturbations. We consider three different types of disturbances based on topological and dynamical perturbations. Topological perturbations are static removals of nodes that might mimic either random disruptions or targeted attacks\cite{albert2000error}. Dynamical perturbations start with the disruption of a single node, generally the one with highest degree, that triggers a cascade of failures\cite{motter2002cascade,motter2004cascade}.

In random disruptions, a fraction $p_{fail}$ of nodes is chosen uniformly random in the network and removed. In targeted disruptions, the fraction $p_{fail}$ of nodes is chosen accordingly to their ranking with respect to a measure of centrality. Usually, the degree is used, but also the betweenness\cite{freeman1977set} -- quantifying centrality with respect to the communication flow -- and k-coreness\cite{seidman1983network} -- based on the core decomposition of a network and characterizing to which nested shell a node belongs to. It is common to quantify the resilience of a network to such perturbations by observing how the relative size of the largest connected component changes as a function of $1-p_{fail}$, i.e. the fraction of survived nodes. This method allows to quantify if the survived nodes are clustered all together or if they form small disconnected clusters which hinder the network's function.

\begin{figure*}
\centering
\includegraphics[width=16cm]{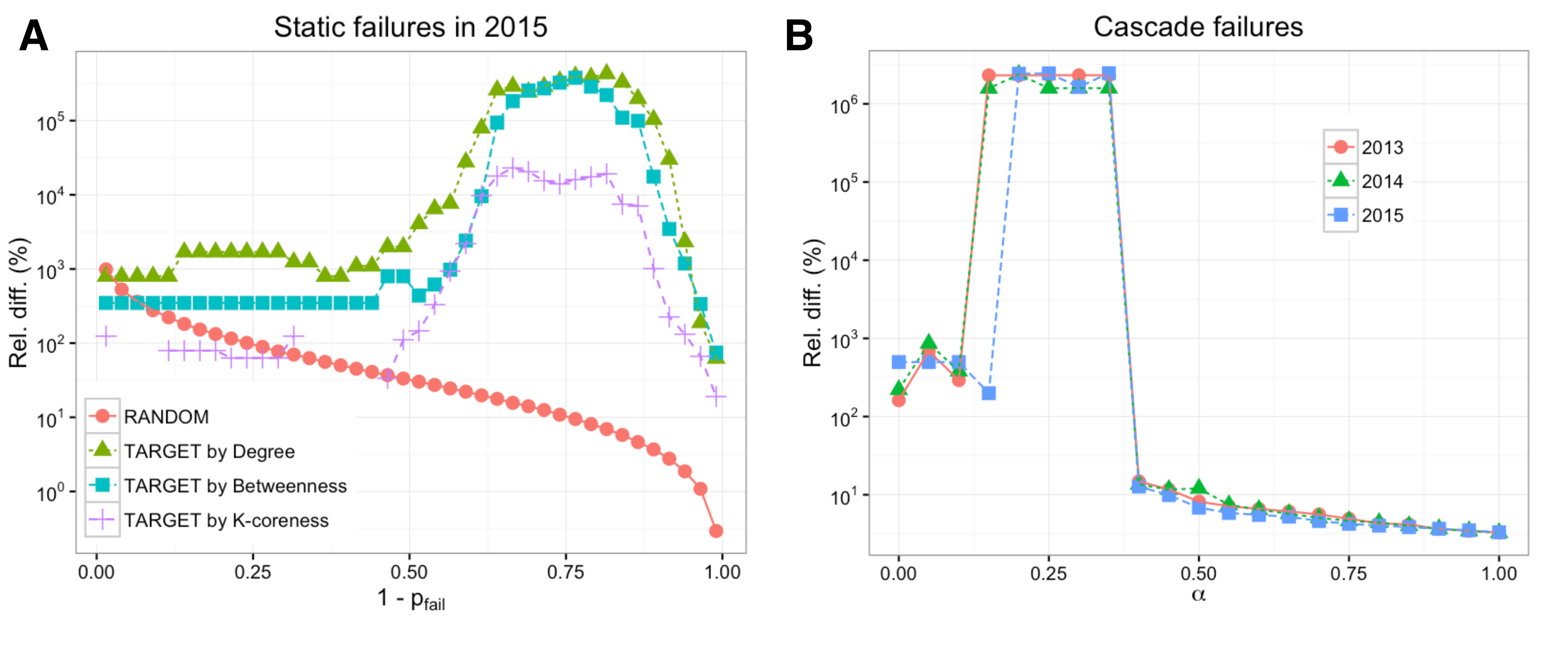}
\caption{{\bf Relative differences in resilience.} ({\bf A}) Differences between the Darknet and the Internet in resilience to random and targeted attacks (see Supplementary Figure~5 for the differences in topological resilience in 2013 and 2014) and ({\bf B}) induced cascade failures.}
\label{fig:resilience-comparison}
\end{figure*}

Non-homogeneous random networks are known to be very robust to random disruptions but very sensitive to targeted attacks. In fact, our findings confirm that both the Internet and the Darknet are fairly robust to random failures, whereas they are more damaged by targeted attacks (see Supplementary Figures 2, 3 and 4). It is worth remarking that the critical point, i.e. the fraction of disruptions for which the size largest connected component of the network is minimum, is very different for the two networks. In fact, while it is enough to target the 10\% of Internet nodes to reach the critical point, in the case of the Darknet much more efforts are needed, requiring a 40\% of disruptions (this result is in excellent agreement with expectation from our model, see Supplementary Figure~6). The corresponding relative differences between the two networks are explicitly shown in Fig.~\ref{fig:resilience-comparison}A, where it is evident that the Darknet is by orders of magnitude, more resilient than the Internet, even with respect to random disruptions.

\subsection{Resilience to dynamical failures}

Another type of disruption, very suitable for communication networks, is based on the inducing cascade failures. The rationale behind this method is that a node $i$ in a communication network is characterized by a certain capacity $\mathcal{C}_{i}$, a fixed feature quantifying the maximum amount of load they can operate with, and a load $\mathcal{L}_{i}(\tau)$, a dynamical feature depending on the state of the network. Nodes with higher degree are assumed to be the ones with higher capacity, and at any time the total load of the network is constant, i.e. $\mathcal{L}=\sum\limits_{i}\mathcal{L}_{i}(\tau)$. If a node with high capacity is disrupted, its load must be redistributed among the other nodes of the network: but if the new loads exceed their capacities, a new set of nodes will suffer a disruption, redistributing the loads through the remaining nodes and so on, thus generating a cascade of failures that can paralyze the system. The dynamics of cascade failures and the resilience of the network can be studied as a function of a parameter $\alpha$ which improves the capacity of each node to $(1+\alpha)\mathcal{C}_{i}$. By varying $\alpha$ and calculating the relative size of the largest connected component at the end of the cascade, we can estimate the required enhancement in capacity to make the network resilient to this attack. The detailed results are shown in Supplementary~Figures 2, 3 and 4, whereas the differences are shown in Fig.~\ref{fig:resilience-comparison}B. Again, the Darknet is much more resilient than the Internet to this catastrophic cascade of failures, requiring just $\alpha\approx 0.2$ to remain fully operative, whereas the Internet requires at least $\alpha=0.28$ to keep operative almost the 90\% of its nodes (full operations is guaranteed for values of $\alpha$ close to 1). This result is, one more time, in excellent agreement with expectation from our model (see Supplementary Figure~6). The relative differences between the resilience of the two networks clearly indicate that before and close to the critical point the Darknet is more resilient than the Internet. This property has direct economic implications, because larger the value of $\alpha$ higher the costs to make the network more robust.

\section{Conclusions and Discussion}

We have investigated the structural properties of the Darknet, the communication network developed in the last two decades to guarantee safe and anonymous navigation. The Darknet exhibits some interesting features that are not shared by the structure of the Internet. Triadic closure in the Darknet is more likely than in the Internet, with communication paths much shorter in the former. Like the Internet, the Darknet is characterized by a non-homogenous connectivity distribution and the presence of higher-order degree-degree correlations. However, the topology of the Darknet is more interesting because of the peculiar heavy-tailed scaling of the degree distribution, with scaling exponent close to -1 and cutoff, at variance with the Internet, appearing more like a power-law with scaling exponent close to -2 and no evident cutoff.

The rich-club analysis has revealed the lack of a core of highly central nodes interconnected each other, at variance with the Internet where this effect is remarkable. We argue that such topological differences are responsible for the different resilience exhibited by the two communication systems in response to random disruption, target attacks and induced cascade failures. We have thoroughly shown that the peculiar topology of the Darknet, characterized by highly clustered communication circuits, small characteristic distance between hops and lack of a rich core, makes this network much more resilient than the Internet as a result of adaptive changes in response to the attempts of dismantling it across time.

While the resilience of the Internet is not significantly changing over time, the resilience of the Darknet is still changing. In fact, its resilience to topological disruption slightly decreases between 2013 and 2014, remaining unchanged in 2015, whereas in the same year the Darknet become slightly less resilient to induced cascade failures. Together with the trends revealed by other structural descriptors, such as decreasing clustering, slightly increasing characteristic length and increasing assortativity, we argue that the Darknet might be undergoing a transition from decentralization to centralization of its services. It will be interesting to confirm this prediction in the future, when longer historical data will be available.

By mimicking how the Darknet actually works, we have proposed a model -- based on a preferential attachment mechanism depending on exogenous properties such as aging and bandwidth and independent on endogenous properties such as node's degree -- to reproduce the empirical degree distribution with remarkable accuracy. The analysis shows that our model is sufficient to understand the structural correlations and the robustness of the Darknet.

We recognize that there are many possible flaws in the representation of both systems, the Darknet and the Internet, as to claim to have the true topological networks. Nevertheless, from the data available it is indeed possible to start thinking about properties of the structure and its implications. The comparison between both structures has been developed at a descriptive level from the underlying graphs, with no other goal than using the Internet as a benchmark to better highlight some salient features of the Darknet. Finally, the proposal of a model driven by the main features of the Darknet seems to be enough to capture its more prominent topological features.

Summing up, the main result of this work is the indication that the mechanisms adopted to guarantee anonymous traffic in the Darknet and to improve the cyber-security of its users could be the main responsible for the peculiar topology of the Darknet observed so far, and its resilience.

\begin{acknowledgments}
The authors thank Robert Annessi and Martin Schmiedecker for providing the Tor raw data and support on data processing. MDD acknowledges financial support from the Spanish program Juan de la Cierva (IJCI-2014-20225). AA acknowledges financial support from ICREA Academia and James S.\ McDonnell Foundation and Spanish MINECO FIS2015-71582.
\end{acknowledgments}

\bibliographystyle{apsrev4-1}
\bibliography{darknet}

\end{document}